# The effect of incident angle of pumping light on Cholesteric Liquid Crystal tunable laser wavelength

Xiangbao Yin[12], Yongjun Liu[1], Lingli Zhang [3], Zhangmin Wu[1], Bofan Huo[2]

1 Harbin Engineering University, Harbin, China

2 Heilongjiang University of Science and Technology, Harbin, China

3 Harbin Institute of Technology, Harbin, China

**Abstract—One notable feature of dye doped cholesteric liquid crystal is the variation of pitch and refractive index as the incidence angle of the pumping light shifts. Based on this feature, this paper studies the effects of pumping light's incidence angle on emission properties of the dye doped cholesteric liquid crystal tunable laser. First, we investigated the relationship among the concentration of chiral reagent, the incidence angle of the pumping light, and the pitch of liquid-crystal display (LCD); then we made a tunable-wavelength laser and pumped the prepared sample with 532nm laser outputted from the Nd: YAG multi-frequency pulse laser. As the incident angle of the pumping light shifts between 20 ° ~ 90 °, the tuning range of the wavelength emitted by the laser reaches 10.73nm, ranging between 647.38nm and 658.11nm.**

## I. INTRODUCTION

As early as in 1998, it was reported for the first time that mirrorless lasing was realized with cholesteric liquid crystal(CLC) by V. I. Kopp et al.[1]. in recent years researchers gradually increased,Yuhua Huang[2] vertically put one side of the dye-doped CLC cell near a thermal platform of 40 ℃ , forming a temperature gradient in the cell. Since the temperature would influence pitch, the wavelengths of lasers were tunning from 577nm to 670nm by pumped different position of the cell. L. M. Blino from Russian Academy of Sciences and GciPParrone et al.[3] from Calabria University of Italy adopted a light-tight hollow electrode to replace a ITO electrode in traditional liquid crystal cell, as a lkHz square-wave voltage being applied on two electrodes of the liquid crystal cell, 25nm wavelength tuning range (600nm to 625nm) was obtained. Moreover, some research [4-5] also dropped the mixture of liquid crystal and dye into the pre-made photonic crystal, and obtained tunable laser output with low threshold value.

By summarizing the above mentioned research results, most of the researchers took the influence of temperature, doping concentration of the dye, pressure, etc. On the thread pitch of liquid crystal as the tuning methods, i.e. influencing the optical band gap formed by CLC to reach the target of tuning output laser wavelength. In this article, we studied the tuning laser by varying the incident angle of pumping light.

## II. EXPERIMENT

### *A. Preparation of laser sample*

Glass plate with a thickness of 1.1mm was selected as the base plate of laser sample,Evenly apply polyimide (PI) alignment layer on one side of the base plate by spin coating, after treatment of rubbing alignment, make empty cells by arranging the friction direction of two base plates as anti-parallel, and control the thickness of liquid crystal layer by utilizing spacers [6]. The cell gap was 10 μ m. Nematic LC BHR33200 ( $n_o$=1.508, $n_e$=1.657, $\Delta n$ =0.149 at 20°C, 589nm;clearing point is 65°C); left-handed chiral additive S811 is selected as the chiral additive，supplied by Beijing Bayi

Space LCD Materials Technology Co . Ltd.dye4-Dicyanomethylene-2-methyl-6-(4-methylaminostyryl)-4H-pyrane (DCM) is elected as the laser dye, supplied by American Exciton Company; and UV-2450 UV spectrophotometer made by SHIMUZU of Japan is adopted for transmission spectrum test.The material was a mixture of 72% nematic LC BHR33200,27% left-handed chiral additive S811 and 1% laser dye DCM, stir well by ultrasonic oscillator after mixing, and inject into the prepared empty sample cells.

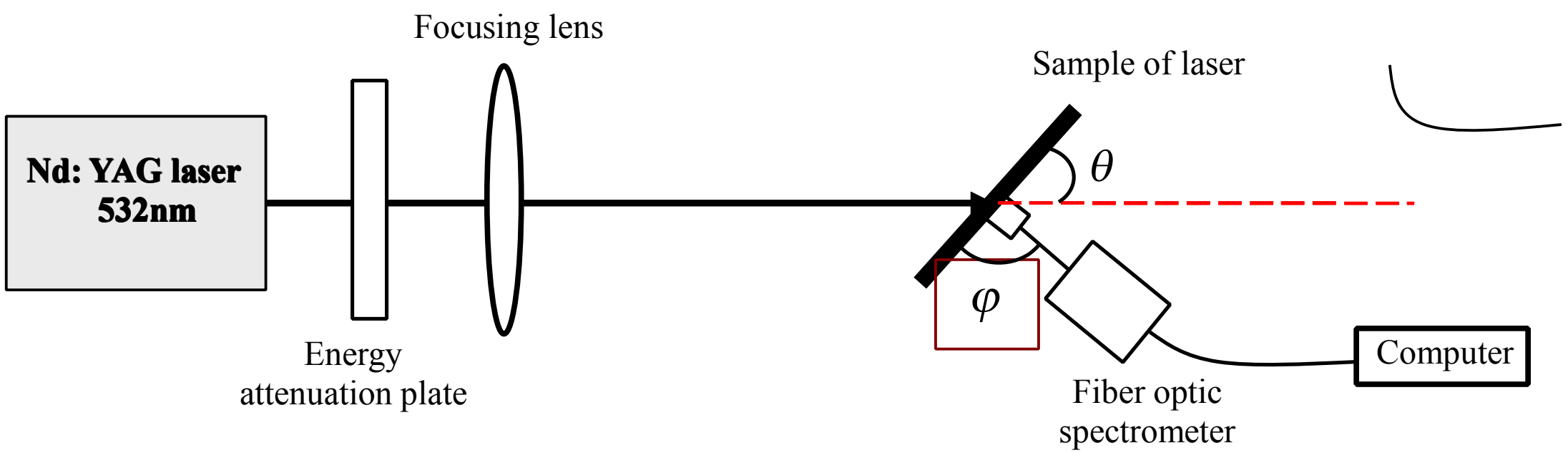

Figure1 Experimental diagram of the dye doped CLC laser

*B. Emission characteristics experiment of tunable laser varying with the incident angle of pumping light*

By mixing nematic LC and chiral additive, CLC can be formed, which has a self-organized periodic helix structure, i.e. periodic reflective index modulation, can be regarded as one-dimensional photonic crystal. If fluorescence spectrum of the dye overlaps photonic band gap of the CLC, narrow linewidth laser can be produced under the excitation of pump light. Adopt Dawa-100Nd: YAG frequency doubling pulsed laser provided by Beijing Beamtech Optronics Co, Ltd., with a repetition frequency of 1Hz. Pumping laser is focused on the sample by convergent lens after passing energy attenuation plate, there is a specified angle between the face normal of the sample and the pumping light, area of the light spot is about $2mm^2$ , which will be imported into computer for processing from the spectrometer by emission spectrum received by the

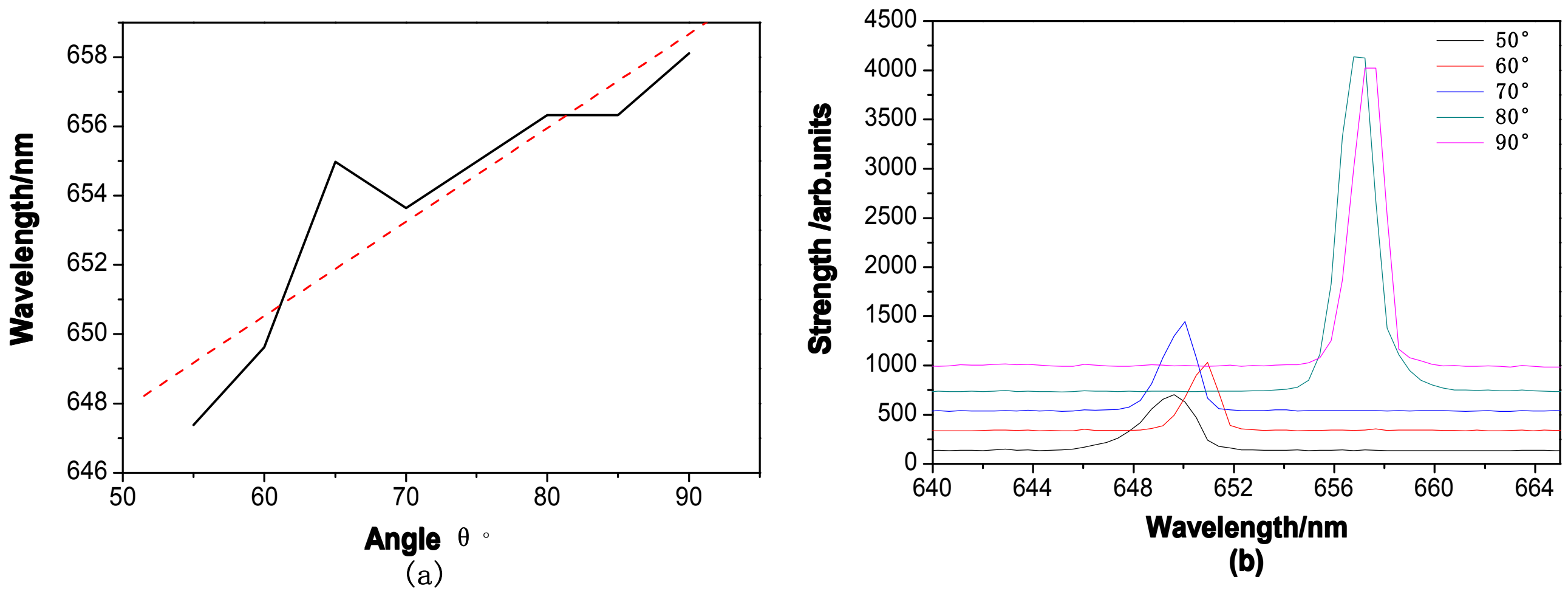

Figure2 Emission wavelength as a function of incidence angle $\theta$

fiber optic probe perpendicular to the surface of the sample. The schematic diagram of experimental optical path is shown in Figure1. The emission laser will be produced perpendicular to the surface of the sample. With angle θ between the pumping light and the face normal of the laser sample changing from 90° to 55°, effective refractive index of the LC has corresponding variation, which resulted in the variation of the cavity length of the laser resonant cavity. And wavelength of emission lasers shifted.

## III. Experiment results and analysis

The emission lasers of the sample with liquid crystal layer thickness of 10μm， were studied by changing the angle θ between the pumping light and the surface of the laser sample. The results were shown in Figure 2. Figure 3(a)shows that the wavelength changes with the angle of θ and Figure 2(b)shows the tranmission intensity respectively when θ =20° 、30° 、40° and50° .It shows that wavelength of emission peak appears at 658.11nm when θ=90°, and shifts to the direction of shortwave with decrease of the angle when θ=55°, the shortest wavelength of 647.38nm will appear .The turning range of whe wavelength is from 647.38 to 658.11nm and reaches 10.73nm. Please see Figure 3 for transmission spectrum of the CLC. The factors that results in gradual decrease of wavelength is: since liquid crystal is a sort of material with high bi-refringence, the refractive index of incident light can be periodically modulated by the periodic helix structure of CLC. The intensive Bragg reflection appears for incident light on the surface of CLC, when incident light enters vertically, central wavelength of Bragg reflection is $\lambda_0 = n \cdot p_0$ [7], in which po is the length of CLC pitch, $n = (n_o + n_e)/2$ is the average refractive index of liquid crystal, central wavelength of Bragg reflection photonic bandgap is determined by Formula $\lambda_0 = n \cdot p_0 \cos\Theta$ in which $\Theta = \sin^{-1}\left[\frac{1}{n}\sin\left(\frac{\pi}{2} - \theta\right)\right]$. From the above-mentioned formula, we can find that reflection photonic bandgap of cholesteric liquid crystal CLC gradually shifts to the direction of short wavelength with the decrease of θ, appearing “blue shift”.

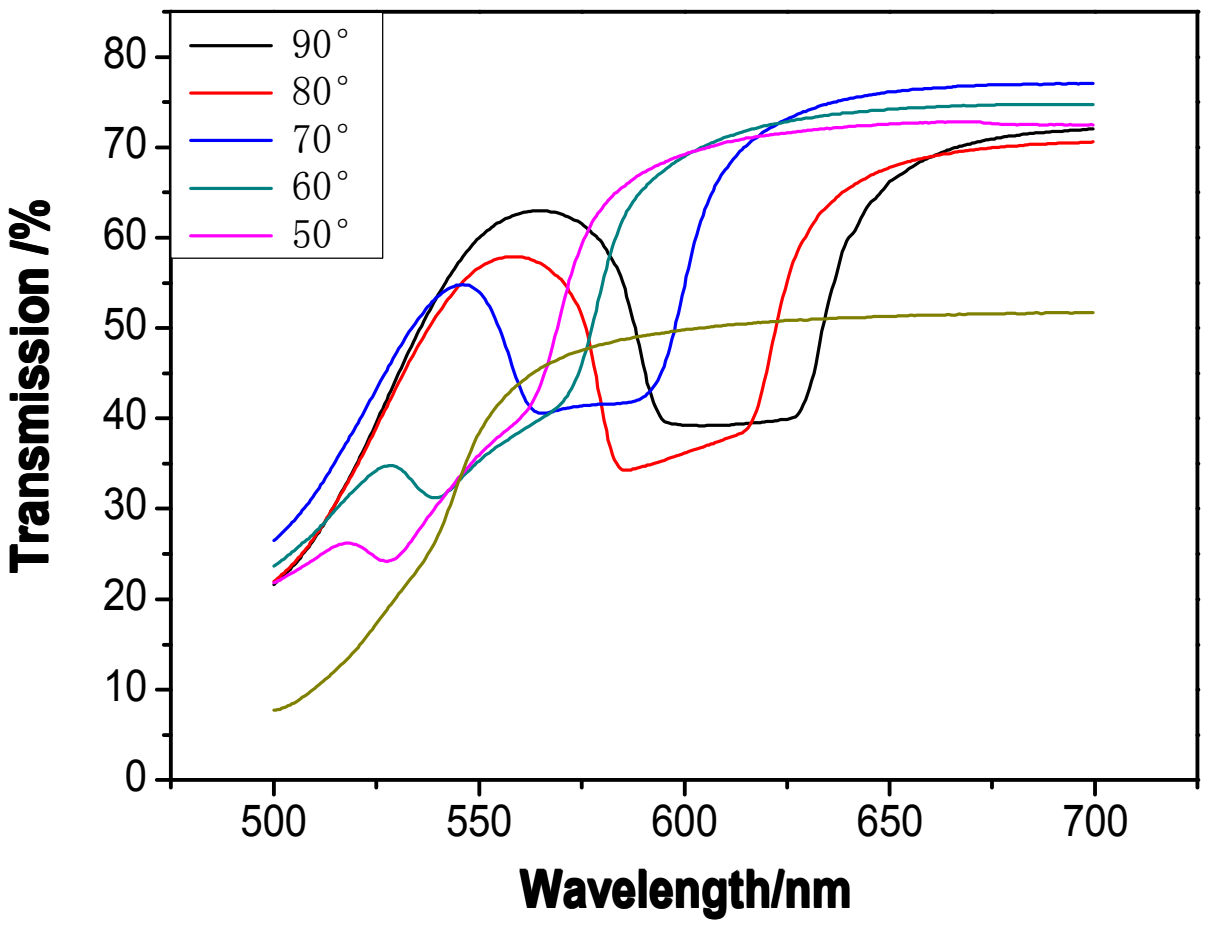

Figure3 the transmission spectrum of CLC

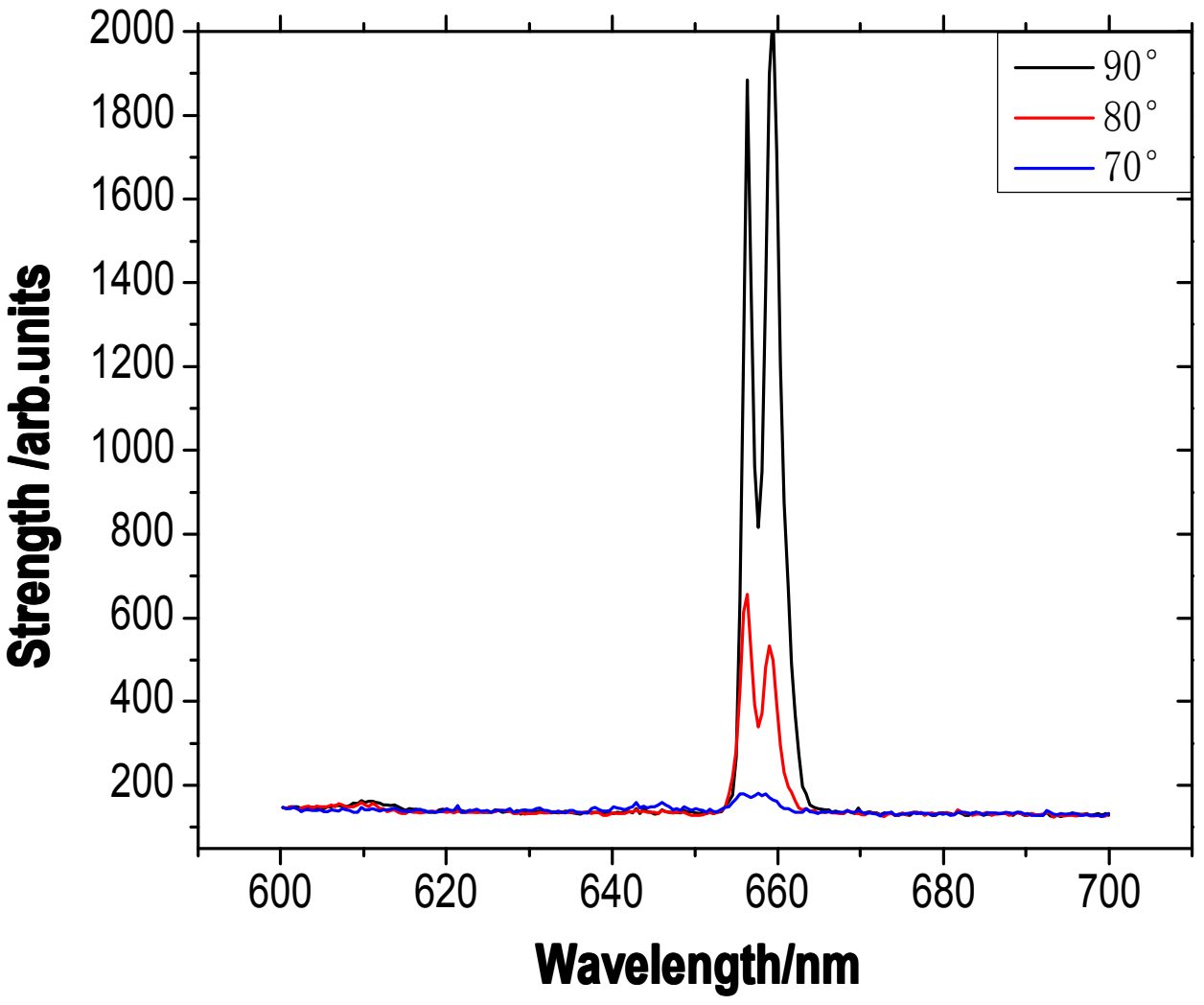

Figure4 Emission spectrum of the dye doped CLC laser at different φ

To avoid the impact of pumping light on experimental, select Angle was θ=75°, the wavelength of emission light was studied by varying the angle φ between the probe of fiber spectrometer and the surface of laser sample,Wavelengths of emission light under different φ were given in Figure 4.The result showed that the wavelength of emission light would not change under different φ.

## IV. CONCLUSION

Tunable laser can be constructed by utilizing CLC and dye DCM. Regulating incident angle of the laser of CLC, tunable effect of emission wavelength can be obtained. The tuning range of laser emission wavelength with the incident angle of pumping light is from 647.38 to 658.11nm, and reaches 10.73nm. The structure of CLC is easier to realize, more convenient to control, having a good application prospect.In the work, it was found that at a given θ, control angle φ between the probe of fiber optic spectrometer and the surface of laser sample, the reason which the wavelength of emission light does not vary at different φ need to be further researched. Next, the work that decreases the threshold value of laser emission, improves the slope efficiency of laser emission shall be carried out.

## ACKNOWLEDGEMENT

This work was supported by the National Natural Science Fund (Approval Number: 61107059, 61077047), China Postdoctoral Science Foundation (Approval Number: 2012M510921) and Heilongjiang Province Postdoctoral Science Foundation (Approval Number: LBH-Z10216)